\documentclass{aa}  

\usepackage{graphicx}
\usepackage{txfonts}
\begin{document}

   \title{Multiwavelength flaring activity of PKS1510-089}

\author{Pedro P. B. Beaklini
          \inst{1},
          T\^{a}nia P. Dominici\inst{2}
          \and
          Zulema Abraham\inst{1}
          }

   \institute{Instituto de Astronomia, Geof\'{\i}sica e Ci\^{e}ncias Atmosf\'{e}ricas,
              Universidade de S\~{a}o Paulo. Rua do Mat\~{a}o 1226, 05508-090, S\~{a}o Paulo/SP, Brazil.\\
              \email{pedro.beaklini@iag.usp.br}
         \and
             Museu de Astronomia e Ci\^{e}ncias Afins, Minist\'{e}rio da Ci\^{e}ncia, Tecnologia, Inova\c{c}\~{o}es e Comunica\c{c}\~{o}es (MAST/MCTIC), Rua General Bruce 586, 20921-030, Bairro Imperial de S\~{a}o Crist\'{o}v\~{a}o, Rio de Janeiro, Brazil\\
             }

   \date{Received xxxxxx xx, xxxx; accepted xxxxx xx, xxxx}
\titlerunning{Flaring activity of PKS1510-089}
\authorrunning{P.P.B. Beaklini, T. P. Dominici, Z. Abraham}

 
  \abstract
   {}
   {In this work, we analyse the multiwavelength brightness variations and flaring activity of  FSRQ PKS1510-089, aiming to constrain the position of the emission sources.}
   {We report 7 mm (43 GHz) radio and {\it R}-band polarimetric observations of PKS1510-089. The radio observations were performed at the Itapetinga  Radio Observatory, while the polarimetric data were obtained at the Pico dos Dias Observatory. 
The 7 mm  observations cover the period between 2011 and 2013, while the optical polarimetric observations were made between 2009 and 2012.} 
   {At 7 mm, we detected a correlation between four radio and $\gamma$-ray flares with a delay of about 54 days between them; the higher frequency counterpart occurred first. Using optical polarimetry, we  detected a large variation in polarization angle ({\it PA}) within two days associated with the beginning of a $\gamma$-ray flare.
Complementing our data with other data obtained in the literature, we show that {\it PA} presented rotations associated with the occurrence of flares.}
   {Our results can be explained by a shock-in-jet model, in which a new component is formed in the compact core producing an optical and/or $\gamma$-ray flare,  propagates along the jet, and after some time becomes optically thin and is detected as a flare at radio frequencies.  The variability in the polarimetric parameters can also be reproduced; we can explain  large variation in both {\it PA} and polarization degree ({\it PD}), in only  one of them, or in neither, depending on the differences in {\it PA} and {\it PD} between the jet and the new component.}


   \maketitle
%

\section{Introduction}

\label{int}

Blazars are active galactic nuclei (AGNs) characterized by non-thermal spectral energy distribution (SED), polarized emission, and  a high level of variability, from radio frequencies to $\gamma$ rays, sometimes on very short timescales  \citep{fos97,ghi98,aha07}.
These extreme properties are believed to be a consequence of the very small angles between the relativistic jets, where the emission occurs, and the line of sight \citep{GUI89,urr95,sam96,mar03,gui08}.

The SED of blazars presents two broad features, a  low-frequency component attributed to synchrotron radiation from relativistic electrons and a high-energy component that could be produced by different processes involving leptons and/or hadrons \citep{bot10}. In the leptonic models, the high-energy emission is the result of inverse Compton (IC) interaction between the same electron/positron distribution that produces the low-energy component of the SED and low-energy photons, produced  in the synchrotron process (SSC), the accretion disk, the broad line region (BLR), or the dusty torus (DT). In the hadronic models, the high-energy component of the SED is due to synchrotron radiation of the proton population, while at TeV energies, synchrotron radiation of pions and muons could predominate \citep{dil15}.
Both leptonic and hadronic models were able to reproduce snapshots of the SED of blazars detected by Fermi/LAT  at high energies (HE), assuming equilibrium between high-energy particle injection, energy losses, and escape from the emission zone and conditions close  to equipartition between electron/proton energies and magnetic fields, but the SEDs of those objects for which very high energy (VHE) was detected with Cherenkov arrays can  be fitted better using hadronic models \citep{bot13}.

Considering the variability detected in the HE emission, very short doubling timescales imply  a very small size for the emission region, favouring its position at the base of the relativistic jet;  the low-energy photons involved in the external Compton (EC) process could then be located in the BLR \citep{ghi10,nal12}. In that case, the high-energy $\gamma$-ray photons would be absorbed by the dense plasma via pair production, producing a high-energy  cut-off in the SED \citep{pou10}.  This cut-off is in fact seen in most of the flat spectrum radio quasars (FSRQs), where strong and broad optical emission lines reveal the existence of BLR clouds \citep{abd10b}, except in a few of them where VHE emission is also detected.

Multiwavelength variability and polarization properties make the location of the high-energy emission even more uncertain. In general, there seems to be a delay between variability at different frequencies;  the high-frequency emission occurs first \citep{bot88,ste94,ste98,cha08,lar08,pus10,bea14,fur14,max14}, as is expected from optical depth considerations if the emission region expands as it moves outward from the base of the relativistic jet. 

On the other hand, detection of simultaneous radio and $\gamma$-ray flares  together with large variations in the optical polarization angle ({\it PA})  \citep{mar08,mar10,jor10,agu11,lar16} could imply that the IC emission  originates downstream of the jet. In that case, the small sizes 
required by short timescale variability were explained as interaction between small turbulent  regions 
and a standing shock \citep{mar14,kie16}.

\begin{table*}                                                                                                                                                          
        {\small                                                                                                                                                 
        \caption{Radio  flux    density obtained in this work}
        \hfill{}                                                                                                                                                        
        \begin{tabular}{        c       c       c       c       c       c       c       c}                                                                                      
        \hline\hline                                                                    
        Date & JD-2400000 & Flux &      Error & Day     & JD-2400000 & Flux & Error\\
     & & \, \, \, Density (Jy) & &      & &   \, \, \, Density (Jy) & \\
\hline 
2011-01-24 & 5585 & 1.84 & 0.39 & 2012-04-23 & 6040 & 8.18 & 0.99 \\
2011-01-29 & 5590 & 2.68 & 0.42 & 2012-05-17 & 6064 & 6.29 & 0.38 \\ 
2011-03-24 & 5644 & 2.19 & 0.37 & 2012-05-20 & 6067 & 6.26 & 0.35 \\ 
2011-03-30 & 5650 & 3.42 & 0.53 & 2012-06-13 & 6091 & 5.83 & 0.65 \\ 
2011-03-31 & 5651 & 3.27 & 0.35 & 2012-06-20 & 6097 & 5.82 & 1.04 \\ 
2011-04-01 & 5652 & 2.56 & 0.44 & 2012-07-26 & 6134 & 2.97 & 0.27 \\ 
2011-04-30 & 5681 & 2.44 & 0.61 & 2012-07-28 & 6136 & 5.04 & 0.43 \\ 
2011-05-01 & 5682 & 1.92 & 0.47 & 2012-08-13 & 6152 & 2.94 & 0.33 \\ 
2011-05-31 & 5712 & 3.13 & 0.43 & 2012-08-15 & 6154 & 2.69 & 0.35 \\ 
2011-06-02 & 5714 & 2.24 & 0.38 & 2012-10-30 & 6230 & 3.08 & 0.48 \\ 
2011-07-12 & 5754 & 3.04 & 0.34 & 2012-11-28 & 6259 & 3.24 & 0.81 \\ 
2011-07-27 & 5771 & 2.42 & 0.32 & 2012-11-30 & 6261 & 5.00 & 0.34 \\ 
2011-08-28 & 5801 & 5.09 & 0.35 & 2013-01-24 & 6316 & 4.22 & 0.66 \\ 
2011-08-30 & 5803 & 3.67 & 0.35 & 2013-02-26 & 6349 & 3.57 & 0.62 \\ 
2011-09-26 & 5830 & 4.62 & 0.32 & 2013-04-04 & 6386 & 3.60 & 0.53 \\ 
2011-09-29 & 5833 & 4.64 & 0.44 & 2013-04-10 & 6392 & 2.94 & 0.31 \\ 
2011-10-28 & 5862 & 4.81 & 0.40 & 2013-05-11 & 6423 & 3.96 & 0.44 \\ 
2011-11-26 & 5891 & 5.81 & 0.46 & 2013-05-13 & 6425 & 2.23 & 0.52 \\ 
2011-11-29 & 5894 & 6.28 & 0.91 & 2013-07-17 & 6490 & 2.48 & 0.32 \\ 
2011-12-22 & 5917 & 5.79 & 0.34 & 2013-08-19 & 6523 & 2.50 & 0.37 \\ 
2012-02-07 & 5964 & 4.64 & 0.34 & 2013-09-02 & 6537 & 1.76 & 0.41 \\ 
2012-02-11 & 5968 & 5.71 & 0.59 & 2013-09-05 & 6540 & 3.46 & 0.36 \\
2012-03-14 & 6000 & 5.91 & 0.71 & 2013-09-13 & 6548 & 1.84 & 0.31 \\ 
2012-03-21 & 6007 & 6.84 & 0.50 & 2013-09-14 & 6549 & 1.51 & 0.26 \\ 
2012-04-20 & 6037 & 7.09 & 0.95 & & & & \\     
\hline
\end{tabular}}
\hfill{}
\label{radioPKS}
\end{table*}

The blazar PKS 1510-089, object of our study, presents most of the characteristics discussed above.  With redshift $z = 0.360 \pm 0.002$ \citep{tho90}, it is one of the FSRQs that presents VHE emission, a high degree of polarization, large and fast variations in the polarization angle, and simultaneous radio and $\gamma$-ray flares. It is a core-dominated source with typical flux density at radio frequencies in the range of 1 to 4 Jy \citep{ter05,alg11}; its relativistic jet forms an angle of about 3 degrees with the line of sight  \citep{hom02}.

PKS 1510-089 has been the target of several optical monitoring programmes \citep[e.g.][]{vil97,rai98,rom02} and seems to be characterized by periods of quiescence followed by large amplitude flares on timescales of several days to months.  At $\gamma$ rays, it was discovered as a HE source by EGRET \citep{har99}, but the existence of significant brightness variability at GeVs has only been verified since 2008, based on AGILE \citep{dam09} and Fermi/LAT observations \citep{abd10b}, while the first detection of the source at TeV energies occurred in 2009, by H.E.S.S. telescopes \citep{abr13}.

Multiwavelength campaigns were organized during HE activity \citep{ori13,ale14,magic16,cas16}
and monitoring at radio frequencies are performed on a regular basis by different programmes, e.g. F-GAMMA (FERMI-GST AGN Multi-Frequency Monitoring Alliance), GASP (GLAST-AGILE Support Program), UMRAD (University of Michigan Radio  Observatory), OVRO (Owens Valley Radio Observatory) 40-meter telescope monitoring at 15 GHz. 

It has long been known that  PKS1510-089 is a highly
 polarized source at optical wavelengths \citep{app67,moo84}. At radio wavelengths, VLBA maps show the existence of a highly polarized and variable core \citep{jor05,jor07,lin11}, variability that seems to be correlated with gamma-ray events \citep{mar10,sas11,ori13}.

In 2011, the $\gamma$-ray light curve obtained from Fermi/LAT observations showed the existence of successive  flares  \citep{fos13,ale14}, while the radio flux started a continuous increase some weeks after the first $\gamma$-ray flare \citep{nes11,ori11,bea11a,ori13}. At $R$ band, the light curve obtained by \citet{smarts} using the SMARTS telescopes (Small and Moderate Aperture Research Telescope System) did not show outbursts simultaneous with the $\gamma$-ray flares \citep{smarts}. However, the existence of many gaps during this period could hide the existence of any correlation.

\citet{mar10} claimed the detection of a rotation higher than $700^{\rm o}$ in {\it PA} at $R$ band before the occurrence of a $\gamma$-ray flare observed by Fermi/LAT \citep{abd10c}, presenting a similar behaviour to that detected in BL Lac \citep{mar08}. The authors interpreted it as a consequence of the existence of a helical magnetic field in the path of a new jet component before it crosses an optically thick core, located far from the central engine. However, the detection of that large continuous rotation is still under debate \citep{sas11,jem16}.

 \begin{figure*}
        \center
        \includegraphics[width=16cm]{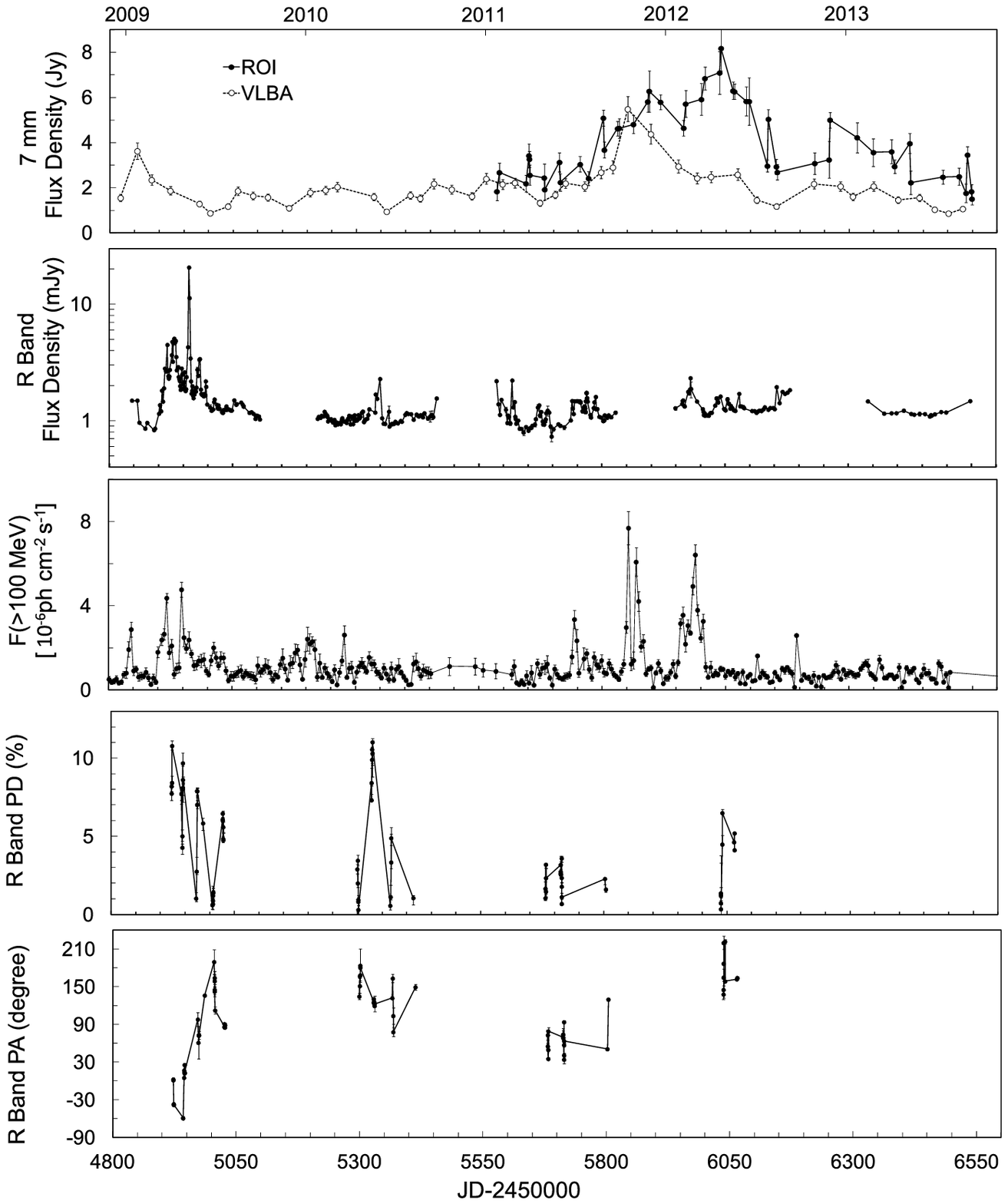}
        \caption{From top to bottom: 7 mm radio light curve obtained in this work together with the 7 mm peak flux density of VLBA images obtained from the VLBI-BU BLAZAR monitoring programme; $R$-band photometry from SMARTS \citep{smarts}; Fermi/LAT $\gamma$-ray light curve for energies $>100$ MeV, binned in five-day intervals \citep{abd09a,abd09b,abd10a};  $R$-band polarization degree and position angle obtained in this work. }
        \label{all}
 \end{figure*}

In 2009 we started a monitoring programme of some bright blazars at $43$ GHz (7 mm) and at optical polarimetry. PKS 1510-089 was included in our optical sample, but not in the radio sample as it was considered a weak source to be observed  with a S/N that was good enough for variability analysis. However, after the detection  of gamma-ray flares in 2011 \citep{sai13,fos13}, we followed the increase in the radio flux density, also detected by other authors \citep{bea11a,bea11b,nes11,ori11}, and then we started to monitor  PKS 1510-089 regularly, 
also at 7 mm. 

In this work we present the results of these observational efforts. The data were compared with those available in the literature at other wavelengths, including gamma-ray observations from Fermi/LAT. Special attention was given to the epochs of intense activity in the form of strong flares.  In section \ref{obs}  we describe the observational programme and the data analysis. The  results from our observations are presented in section \ref{res}, and they are discussed and compared with other data sets available the literature in section \ref{dis}.
Finally, in section \ref{conclusion} we present our conclusions.

\hspace{1.0cm}  


\section{Observations}
\label{obs}

\subsection{Radio observations}
\label{radio}

 The 7 mm observations of PKS 1510-089 began  in January 2011 after the detection of $\gamma$-ray flares by Fermi/LAT, and continued monthly until April 2013.
They were made at the Itapetinga Radio Observatory,  in Atibaia, S\~{a}o Paulo, Brazil. The radiotelescope consists of a $13.7$ m radome enclosed antenna and a room temperature K-band receiver, with a 1 GHz double sideband and a noise temperature of about 700 K, which provides a 2.4 arc min HPBW. For instrumental calibration we used a room temperature load and a noise source of known temperature in a procedure that takes into account radome and atmospheric absorption  \citep{abr92}.  SgrB2 Main, an HII complex close to the galactic centre, was used as flux calibrator.

We used the scan method (on-the-fly) for the observations. Each scan lasted 20 s,  was centred at the source coordinates, and had an amplitude of 30 arc min in elevation or azimuth. As we were observing a point source, we used the scans in both directions to verify the pointing accuracy. During one scan, we measured 81 uniformly spaced points; each observation consisted of 30 scans and the instrumental calibration was performed every 30 minutes.  For SgrB2 Main, which has an angular size larger than the beam width, the scans were made along the right ascension coordinate. On a typical day, we performed between 18 and 20 observations of PKS 1510-089. 



\subsection{Optical polarimetry} 
\label{pol}

Optical polarimetric observations were carried out between 2009 and 2012 using the 0.6 m  Boller \& Chivens IAG/USP telescope at Pico dos Dias Observatory (OPD, Braz\'opolis, Brazil)\footnote{ Operated by the Laborat\'orio Nacional de Astrof\'\i sica (LNA/MCTI) } and an imaging polarimeter, IAGPOL \citep{mag96}, working in linear polarization mode with a standard {\it R}-band filter. The polarimeter consists of a rotatable achromatic half-wave retarder, followed by a calcite Savart plate. This configuration provides two images of each object in the field with orthogonal polarizations, separated by $1$ mm or $25.5$ arcsec at the telescope focal plane. The simultaneous detection of the two beams allows observations under non-photometric conditions and has the advantage that the sky polarization is practically cancelled out. 
The observations were performed over a few consecutive nights on a monthly basis, with several observations during each night.

Throughout the years of monitoring, we used two different CCDs: a $1024 \times 1024$ pixel CCD of $24$ microns/pixel and
 a $2048 \times 2048$ pixel CCD of $13.5$ microns/pixel, both providing a field of view of about $10^{'} \times 10^{'}$ ($0.67^{"}$/pixel and $0.38^{"}$/pixel, respectively). Each polarization measurement was obtained from eight different wave plate positions separated by $22^{\rm o}.5$, consuming a mean total integration time of about 30 minutes, depending on the quality of the night The images were reduced with IRAF\footnote{IRAF is distributed by the National Optical Astronomy Observatory, which  is  operated  by  the  Association  of  Universities  for  Research in  Astronomy,  Inc., 
under  cooperative  agreement  with  the  National Science Foundation} usual routines for bias and flat field corrections. The PCCDPACk package \citep{per00} was used to calculate the polarization, its parameters, and errors based on \citet{ser74a,ser74b}. The conversion of the  polarization position angle {\it PA} to the equatorial system was made using observations of polarized standard stars in each night (HD$298383$, HD$111579,$ and HD$155197$). The lack of instrumental polarization was checked through the observation of unpolarized standard stars (HD$94851$, WD$1620,$ and HD$94851$).

In an imaging polarimeter like IAGPOL, the total flux density can be recovered by adding both polarimetric components. However, this procedure is reliable only for photometric nights. In order to check the quality of the data obtained with this method, we  used  the public light curves from the Yale/SMARTS monitoring programme \citep{smarts}, available from  the webpage of the project\footnote{ http://www.astro.yale.edu/smarts/glast/home.php }.  Comparing coincident observations between our daily averaged data and those of the SMARTS programme, we found a maximum  error of $30 \%$ in the total flux, a value that we assume as the error for of all our flux density  measurements. Obtaining photometric information from polarimetric data, even with large error bars, allows us to compare total flux and polarization simultaneously.   


SMARTS data were also used to compare our polarimetric results with a better sampled optical light curve. Blazars at declinations smaller than 20 degrees from the list of sources monitored by the Fermi telescope have been observed by the Yale/SMARTS consortium since 2008, with a 1.3 m telescope at CTIO and the dual-channel imager ANDICAM \citep{dep03}. By using a  dichroic  to feed  an  optical  CCD  and  an  IR  detector,  the instrument can obtain simultaneous data in the $B$, $V$, $R$, $J$, and $K$ bands.  For this work, we selected only the $R$ band because it is the same used in our polarimetric observations. The Yale/SMARTS observation of each object are obtained at least once every three days. To compute the calibrated flux densities used throughout this work, we  considered a galactic extinction of $A_{R} = 0.258$ \citep{sch98}.


\subsection{Fermi data}
\label{fermi}

The Fermi Space Observatory provides  daily light curves of blazars, including PKS 1510-089. The telescope was launched in 2008 to explore the Universe in the energy range of 10 keV to 300 GeV. 
In particular, it carries the Large Area Telescope (LAT) as the main instrument, which allows the observation of the entire sky every three hours. A complete description of the LAT instrument and its operation and can be found in \citet{atw09}. 
In this work, we used the Fermi monitoring data available in the Fermi Light Curve page\footnote{http://fermi.gsfc.nasa.gov/ssc/data/access/lat/msl\_lc/}. 

We chose the largest energy band width (Band 3, $> 100$ MeV) to compare with the radio and the polarimetric light curves. In order to smooth the light curve and to compare it with the long variability trend at radio wavelengths, the $\gamma$-ray data were re-sampled in five-day bins.

\begin{figure}
        \includegraphics[width=9cm]{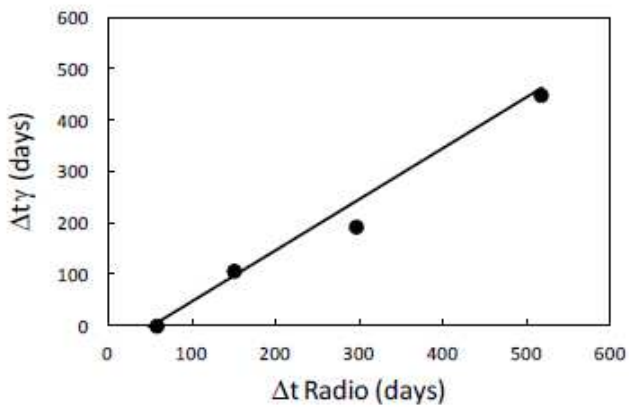}
    \centering
        \caption{Correlation between the time of occurrence of  radio and $\gamma-$ray flares. Zero corresponds to the first $\gamma-$ray flare, on 01 Jul 2011. The dates of the other $\gamma$-ray flares are 24 Oct 2011, 21 Feb 2012, and 23 Sept 2012. The radio peaks correspond to 28 Aug 2011, 29 Nov 2011, 22 Apr 2012, and 30 Nov 2012.}
        \label{new}
\end{figure}
\begin{figure}
        \includegraphics[width=9cm]{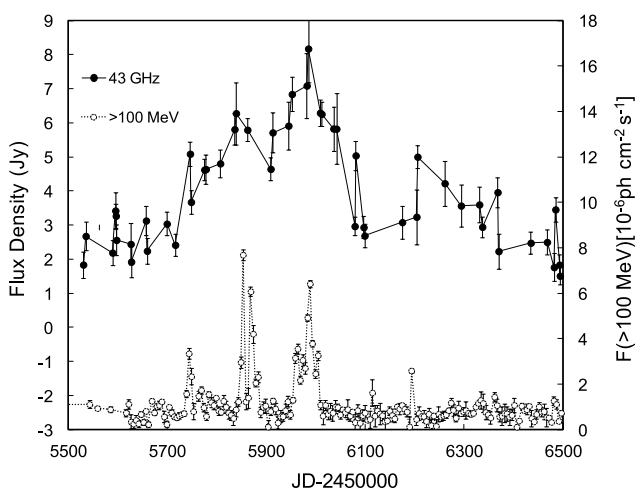}
    \centering
        \caption{Light Curves of PKS 1510-089 at 7 mm (this work, black dots) and at $\gamma$ rays (band 3 data from Fermi/LAT, white dots). The 7 mm light curve is shifted by -54 days.}
        \label{G7}
\end{figure}

\begin{figure}
        \includegraphics[width=9cm]{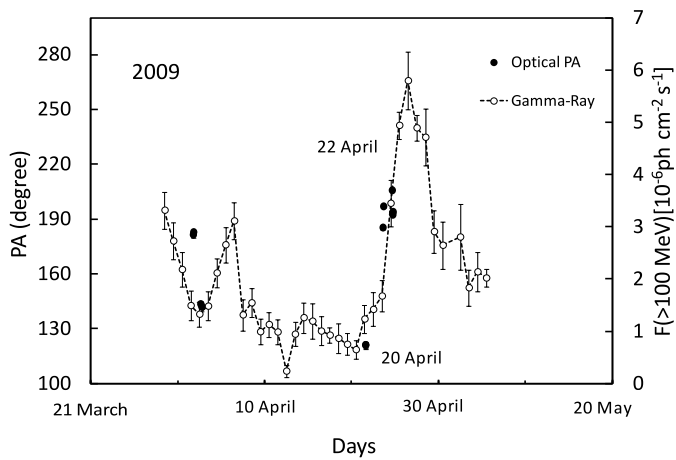}
    \centering
        \caption{Abrupt variation in {\it PA}   simultaneous with the beginning of a $\gamma$-ray flare in PKS 1510-089 during April 2009}.
        \label{Fig2}
\end{figure}

\section{Results}
\label{res}

We present the observed 7 mm flux density of PKS 1510-089 in Table 1 and the {\it R}-band polarimetric results in Table 2. Since there is a $180^\circ$ ambiguity in    {\it PA}, we selected the values  that minimize the difference between consecutive angle measurements, allowing both clockwise and anticlockwise rotations.
In Figure \ref{all} from top to bottom  we show 
the 7 mm Itapetinga light curve together with the total peak intensity of VLBA images obtained from the VLBA-BU-BLAZAR Program\footnote{http://www.bu.edu/blazars/VLBAproject.html}: the {\it R}-band light curve from SMARTS \citep{smarts}, the Fermi/LAT $\gamma$-ray flux at energies higher than 100 MeV \citep{abd09a,abd09b}, and our measurements  of {\it PD} and  {\it PA} at {\it R} band. The $\gamma$-ray data were binned in five-day intervals to smooth the light curve, allowing a better comparison with our 7 mm results.

\subsection{The 7 mm light curve}
\label{7mm}

As can be seen in Figure \ref{all}, the source presented small fluctuations at 7 mm during the first semester of our monitoring programme, with a mean flux density of $(2.6 \pm 0.5)$ Jy, not an unusual level for this object \citep{ter05,alg11}. In the second semester of 2011, about 50 days after the occurrence of a $\gamma$-ray flare detected by Fermi/LAT on 1 Jul 2011 , the brightness started to increase gradually \citep{bea11a,bea11b,nes11,ori11}, reaching the maximum flux density of $(8.2 \pm 0.9)$ Jy, never before detected in this source. It should be noted that this exceptional increase lasted almost one year, although with some  variability, while the decline was faster, taking only two months. Similar behaviour  was found by the F-GAMMA monitoring programme  at frequencies ranging from 2.6 GHz to 142 GHz \citep{ori13} and  in data obtained by the Metsähovi Observatory at 37 GHz \citep{ale14}, including the maximum flux density but taking into account that there is a delay between the different frequencies, as is  discussed in the next section, and that these observations covered a shorter period of time, missing the decaying phase. 
During this period of time, several $\gamma$-ray flares were detected by Fermi/LAT, some of them exceeding the $\gamma$-ray flux of $10^{-5}$ ph cm$^{-3}$ s$^{-1}$, with a doubling time of the order of 20 min \citep{fos13}. 

Comparing our 7 mm light curve with the corresponding VLBA peak intensity data, in the top panel  of Figure \ref{all} we see that the latter, which corresponds to the core emission, reached its maximum value before our single-dish data, implying that after 2012 the jet components included in our single-dish detection contributed considerably  to the total flux density. 

We tried to find a correlation between the $\gamma$-ray and 7 mm light curves, assuming that the continuous increase in flux density was the result of the superposition of radio flares associated with the high-energy events. 
Because of the difference in flare duration at the two frequencies, it was not possible to find a correlation  using  statistical techniques, even those designed for unevenly sampled data as the Discrete Correlation Function \citep{ede88}. Instead, we confirmed the correlation  using the peaks of both light curves. 
For $\gamma$-rays, we choose as peaks the points where the flux density is three times higher than the rms of the quiescent phase; when several flares occurred in a short time interval, we used the mean epoch of the group. 
For radio variability, the  superposition of flares made the identification of the corresponding peaks more difficult; we selected the peaks as the local maxima in the light curve. 
In Figure \ref{new},  the excellent correlations between the flares at the two frequencies can be seen;  the radio flares are delayed by about 54 days relative to the  $\gamma$-ray flares. In Figure \ref{G7} we present the 7 mm and $\gamma$-ray light curves;  the 7 mm data is shifted by -54 days to show the correspondence between the different  flares. 


  \begin{figure*}
        \includegraphics[width=16cm]{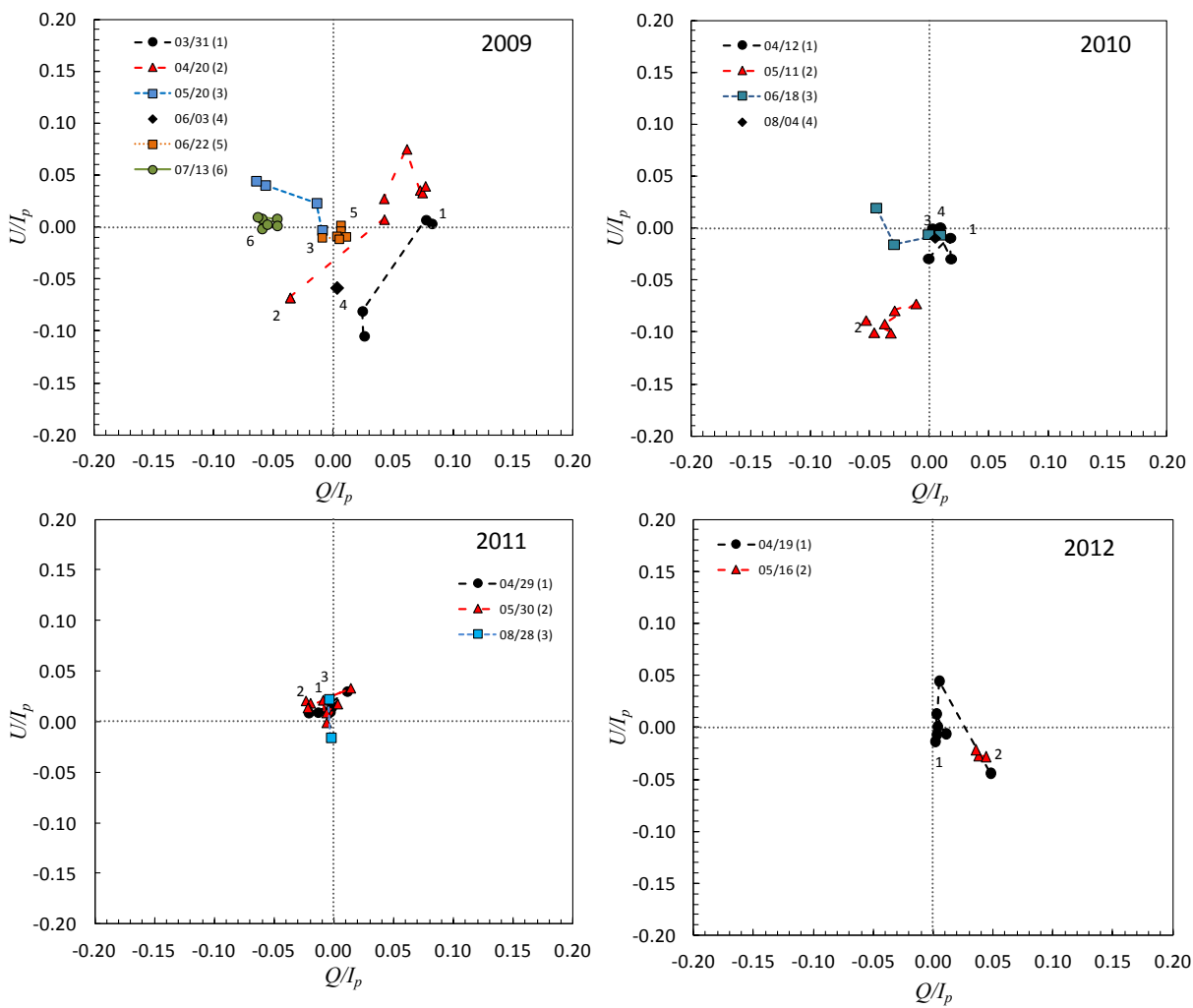}
        \caption{$Q$ x $U$  plane normalized by $I_{p}$ during four different years of our monitoring.}
        \label{QU}
  \end{figure*}
  
  \begin{figure}
        \center
        \includegraphics[width=9cm]{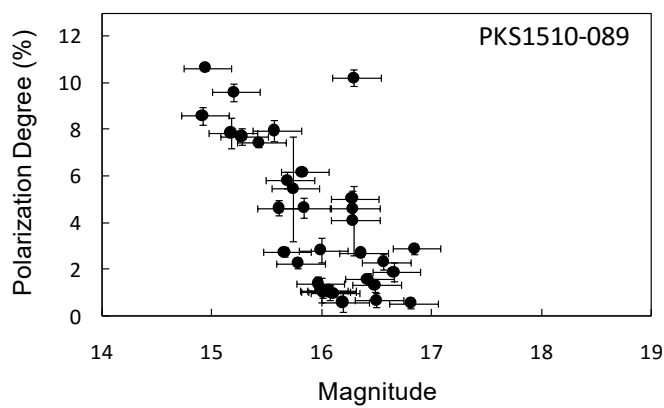}
        \caption{Relation between {\it PD} and  $R$-band magnitude in PKS 1510-089 between 2009 and 2012.}
        \label{polmag}
  \end{figure}


\begin{figure*}
        \includegraphics[width=17cm]{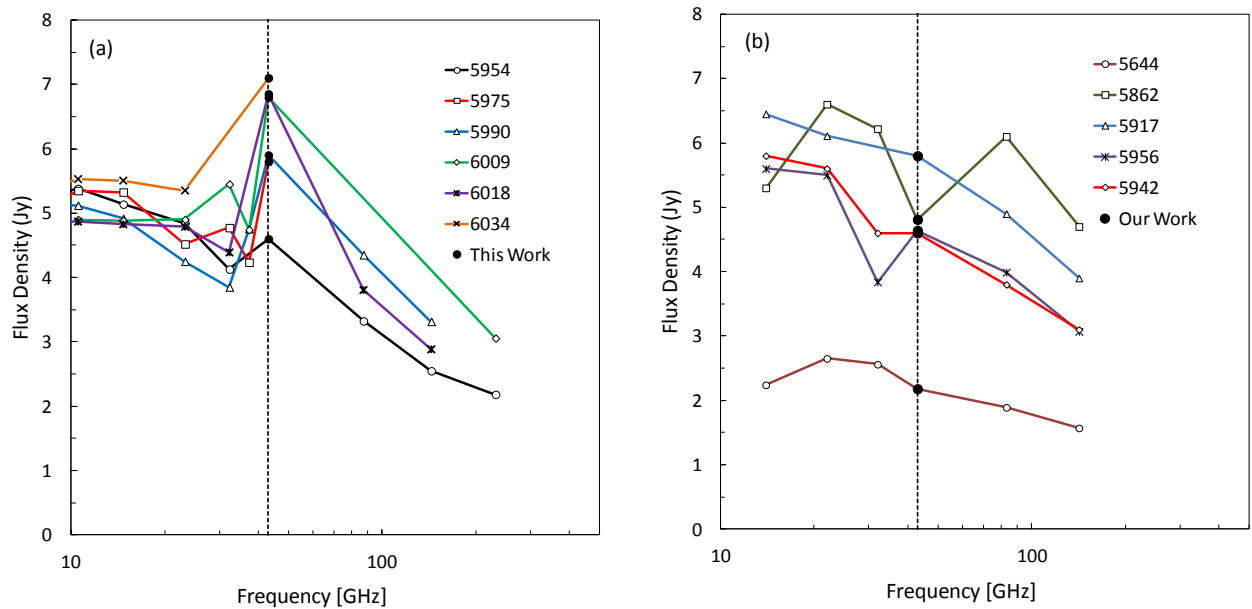}
    \centering
        \caption{Radio spectra of PKS 1510-089 during several epochs with 7 mm data (black dots) obtained in this work superposed with \citet{ale14} observations at (a) and with \citet{ori13} observations at (b).}
        \label{lastt}
\end{figure*}

\subsection{Polarimetry}
\label{pol}

The polarimetric data, shown in the bottom panels of Figure \ref{all}, presented different patterns during 2009, including large variations in {\it PA} and/or {\it PD} in daily timescales and sometimes during the same night.

Between 31 March 2009  and 1 April, five days after the occurrence of a $\gamma$-ray flare and the detection of VHE emission by H.E.S.S \citep{abr13}, there was a change of about $40^\circ$ in {\it PA} in the clockwise direction and a small increase in {\it PD}.  

Between 20 April and 22 April 2009, coincident with the beginning of a $\gamma$-ray flare \citep{abd10c} and the ejection of a new superluminal component \citep{mar10}, {\it PA} changed by $65^\circ$ in the anticlockwise direction and  {\it PD} decrease from 7.7\% to 4.7\%, going back to 8.5\% on 23 April, without any significant change in {\it PA}. The coincidence between the $\gamma$-ray intensity and variations in {\it PA} during this flare can be seen in more detail in Figure \ref{Fig2}.

 From 20 May to 22 May 2009, {\it PD} increased from 1\% to 7.4\% without large changes in {\it PA;}  by the end of June the source was depolarized \footnote{We assumed the boundary definition of 3\% by \citet{ang80} to classify AGNs as polarized.}, going back to 6\% {\it PD} in July.
  
Between 12 April and 14 April 2010 the source was depolarized,  while one month later it reached the maximum measured value of {\it PD}  ($11.02\% \pm 0.05 \%$) after the occurrence of a $\gamma$-ray flare with a doubling timescale of $\sim 0.3$ h \citep{fos13}. 
 
In 2011, when the radio  flux density increased simultaneously with the occurrence of several $\gamma$-ray flares, we found only small amplitude variations in  {\it PD} and {\it PA}, which might have been a consequence of our limited sampling since most of the $\gamma$-ray activity occurred when PKS 1510-089 could not be observed at night. 

In Figure \ref{QU} we show our {\it R}-band  polarization data in the  $Q/I \times U/I$ plane, which  has the advantage of being independent of any criterion used to solve the $180^\circ$ multiplicity. 
Owing to the trigonometric relation between {\it PA} and the Stokes parameters, a continuous rotation of {\it PA} should produce  consecutive changes of quadrants in this plane, as argued by \citet{sas11}. In our data, we detected large changes during or close to the occurrence of flares, and small fluctuations at other epochs. 

We show in  Figure \ref{polmag} the relation of {\it PD} with magnitude  during the polarimetric monitoring.  We found a Pearson coefficient correlation of $r = 0.5$ with the full data set. A positive correlation $r = 0.8$ was obtained by \citet{ike11}, restricted to the observations where the source was brighter than 15.5 magnitudes in {\it V}.

Other polarimetric observations in 2009 at {\it R} band \citep{mar10,jem16}  and  at {\it V} band \citep{sas11} showed a variable behaviour for {\it PD} similar that found in our work, although with even higher maximum values. The behaviour of {\it PA} was also similar, except for the addition of  $180^\circ$  and  the sense of rotation, which was always in the same anticlockwise direction in \citet{mar10}, and in  both directions in \citet{sas11} and \citet{jem16}.


\longtab{

\begin{longtable}{cccccccccccc}

    \caption{ {\it R}-flux density and polarization of PKS 1510-089 obtained in this work \label{151}}\\
         \hline\hline

        Date & JD & Q/I & U/I   & Total Flux & Error & Polarized Flux & Error &  \it{PD} & Error & \it {PA}  & Error \\
     & -2450000 & & &  mJy & mJy  & mJy & mJy & $ \% $ & $ \% $ & degrees & degrees \\
    \hline
    \endfirsthead           
        \caption{continued.}\\
        \hline\hline
    \endhead
    
        2009-03-31  &  4922.81  & 0.0818  & 0.0031  &   1.75  &  0.52  &  0.143 & 0.048  &  8.19  &  0.31   &  1.2     & 1.1 \\ 
        2009-03-31  &  4922.85  & 0.0770  & 0.0058  &   1.75  &  0.52  &  0.135 & 0.050  &  7.73  &  0.58   &  2.6     &  2.1  \\
        2009-04-01  &  4923.66  & 0.0238  & 0.0044  &   2.46  &  0.74  &  0.207 & 0.073  &  8.42  &  0.44  &  -36.8   &  1.5  \\
        2009-04-01  &  4923.79  & 0.0253  & 0.0033  &   2.46  &  0.74  &  0.265 & 0.088  & 10.78  &  0.33  &  -38.2   &  0.9  \\
        & & & & & & & & & & &  \\
        2009-04-20  &  4942.62  & -0.0367  &  0.0044  &  2.3  &  0.69  &  0.177 & 0.063  &  7.71  &  0.44  &  -59.2  &  1.6  \\
        2009-04-20  &  4942.65  & -0.0367  &  0.0025  &  2.3  &  0.69  &  0.177 & 0.059  &  7.71  &  0.25  &  -59.2  &  0.9  \\
        2009-04-22  &  4944.63  &  0.0420  &  0.0024  &  1.7  &  0.50  &  0.072 & 0.025  &  4.27  &  0.24  &  5.1    &  1.6  \\
        2009-04-22  &  4944.70  &  0.0418  &  0.0041  &  1.7  &  0.50  &  0.084 & 0.032  &  5.01  &  0.41  &  16.7   &  2.4  \\
        2009-04-23  &  4945.70  &  0.0609  &  0.0066  &  3.2  &  0.96  &  0.309 & 0.114  &  9.67  &  0.66  &  25.5   &  1.9  \\
        2009-04-23  &  4945.73  &  0.0719  &  0.0041  &  3.2  &  0.96  &  0.257 & 0.090  &  8.02  &  0.41  &  13.2   &  1.5  \\
        2009-04-23  &  4945.76  &  0.0738  &  0.0030  &  3.2  &  0.96  &  0.259 & 0.087  &  8.09  &  0.30  &  12.1   &  1.1  \\
        2009-04-23  &  4945.80  &  0.0764  &  0.0019  &  3.2  &  0.96  &  0.275 & 0.089  &  8.60  &  0.19  &  13.7   &  0.6  \\
        & & & & & & & & & & &  \\
        2009-05-20  &  4972.55  & -0.0098  &  0.0038  &  1.18 &  0.35  &  0.012 & 0.008  &  1.02 &  0.38  &  98.0  &  10.7\\ 
        2009-05-21  &  4973.63  & -0.0145  &  0.0021  &  1.60  & 0.48  &  0.044 & 0.017  &  2.74 &  0.21  &  60.9  &  2.1  \\
        2009-05-22  &  4974.57  & -0.0573  &  0.0019  &  1.99  & 0.60  &  0.139 & 0.046  &  7.01 &  0.19  &  72.4  &  0.8  \\
        2009-05-22  &  4974.60  & -0.0651  &  0.0023  &  1.99  &  0.60  &  0.157 & 0.052 &  7.88  &  0.23 &  72.8  &  0.8  \\
        & & & & & & & & & & &  \\
        2009-06-03  &  4986.57  &  0.0024  &  0.0034  &  1.57  &  0.47  &  0.092 & 0.033  &  5.83  &  0.34  & 136.2  &  1.7  \\
        & & & & & & & & & & &  \\
        2009-06-22  &  5005.54  &  0.0056  &  0.0040  &  1.0  &  0.30  &  0.006 & 0.006  &  0.60  &  0.40  &  189.6  &  19.3  \\
        2009-06-23  &  5006.54  &  0.0056  &  0.0024  &  1.1  &  0.33  &  0.007 & 0.005  &  0.66  &  0.24  &  164.2  &  10.4  \\
        2009-06-23  &  5006.57  &  0.0101  &  0.0046  &  1.1  &  0.33  &  0.015 & 0.010  &  1.36  &  0.46  &  159.1  &  9.6  \\
        2009-06-23  &  5006.60  &  0.0023  &  0.0024  &  1.1  &  0.33  &  0.010 & 0.006  &  0.88  &  0.24  &  142.4  &  7.9  \\
        2009-06-23  &  5006.63  &  0.0042  &  0.0034  &  1.1  &  0.33  &  0.013 & 0.008  &  1.21  &  0.34  &  145.2  &  8.1  \\
        2009-06-24  &  5007.49  & -0.0100  &  0.0027  &  1.2  &  0.36  &  0.017 & 0.008  &  1.41  &  0.27  &  112.4  &  5.5  \\
        & & & & & & & & & & &  \\
        2009-07-13  &  5026.48  &  -0.0602 &  0.0014  &  1.38  &  0.41  &  0.084 & 0.027  &  6.08  &  0.14  &  86.1  &  0.6  \\
        2009-07-13  &  5026.52  &  -0.0601 &  0.0020  &  1.38  &  0.41  &  0.084 & 0.027  &  6.02  &  0.20  &  90.7  &  0.9  \\
        2009-07-13  &  5026.56  &  -0.0638 &  0.0016  &  1.38  &  0.41  &  0.089 & 0.029  &  6.46  &  0.16  &  85.6  &  0.7  \\
        2009-07-14  &  5027.47  &  -0.0476 &  0.0018  &  0.91  &  0.27  &  0.044 & 0.015  &  4.83  &  0.18  &  85.1  &  1.0  \\
        2009-07-14  &  5027.52  &  -0.0558 &  0.0033  &  0.91  &  0.27  &  0.051 & 0.018  &  5.58  &  0.33  &  88.6  &  1.7  \\
        2009-07-14/  &  5027.56  &  -0.0475 &  0.0045  &  0.91  &  0.27  &  0.051 & 0.018  &  4.75  &  0.45  &  89.1  &  2.7  \\
        & & & & & & & & & & &  \\
        2010-04-12  &  5299.73  & -0.0004  &  0.0022  &   0.54  &  0.16  &  0.016 & 0.006  &  2.89  &  0.22  &  134.6  &  2.2  \\
        2010-04-13  &  5300.74  &  0.0178  &  0.0033  &   0.85  &  0.25  &  0.017 & 0.008  &  1.99  &  0.33  &  166.7  &  4.7  \\
        2010-04-13  &  5300.76  &  0.0185  &  0.0130  &   0.85  &  0.25  &  0.029 & 0.009  &  3.44  &  0.04  &  151.2  &  0.3  \\
        2010-04-14  &  5301.72  &  0.0074  &  0.0034  &   0.74  &  0.22  &  0.006 & 0.004  &  0.81  &  0.34  &  168.4  &  12.0 \\
        2010-04-14  &  5301.75  &  0.0093  &  0.0029  &   0.74  &  0.22  &  0.007 & 0.004  &  0.94  &  0.29  &  183.7  &  8.9  \\
        2010-04-14  &  5301.78  &  0.0029  &  0.0027  &   0.74  &  0.22  &  0.002 & 0.003  &  0.29  &  0.27  &  180.6  &  26.7  \\
        & & & & & & & & & & &  \\  
        2010-05-11  &  5328.66  &  -0.0290 &  0.0031  &   2.52  &  0.76  &  0.212 & 0.072 &  8.41   &  0.31  &  124.9  &  1.0  \\
        2010-05-11  &  5328.69  &  -0.0109 &  0.0096  &   2.52  &  0.76  &  0.184 & 0.080 &  7.31   &  0.96  &  130.7  &  3.8  \\
        2010-05-12  &  5329.66  &  -0.0374 &  0.0037  &   0.90  &  0.27  &  0.089 & 0.030 &  9.89   &  0.37  &  123.9  &  1.1  \\
        2010-05-12  &  5329.71  &  -0.0323 &  0.0031  &   0.90  &  0.27  &  0.095 & 0.031 &  10.55  &  0.31  &  126.1  &  0.9  \\
        2010-05-13  &  5330.66  &  -0.0530 &  0.0017  &   3.13  &  0.94  &  0.322 & 0.102 &  10.30  &  0.17  &  119.5  &  0.5  \\
        2010-05-13  &  5330.68  &  -0.0462 &  0.0005  &   3.13  &  0.94  &  0.345 & 0.105 &  11.02  &  0.05  &  122.6  &  0.1  \\
        & & & & & & & & & & &  \\  
        2010-06-18  &  5366.55  &  -0.0006 &  0.0023  &   0.6   &  0.17  &  0.003 & 0.002  &  0.56  &  0.23  &  132.0  &  12  \\
        2010-06-19  &  5367.50  &   0.0093 &  0.0006  &   1.1   &  0.33  &  0.012 & 0.004  &  1.11  &  0.06  &  163.3  &  1.5  \\
        2010-06-20  &  5368.49  &  -0.0296 &  0.0075  &   0.9   &  0.27  &  0.030 & 0.016  &  3.33  &  0.75  &  103.6  &  6.5  \\
        2010-06-20  &  5368.53  &  -0.0445 &  0.0222  &   0.9   &  0.27  &  0.044 & 0.033  &  4.88  &  2.22  &   77.9  &  13.1  \\
        & & & & & & & & & & &  \\   
        2010-08-04  &  5413.46  &  0.0051  &  0.0026  &   1.16  &  0.35  &  0.012 & 0.007  &  1.05  &  0.26  &  149.4  &  7.0  \\ 
        & & & & & & & & & & &  \\  
        2011-04-29  &  5681.63  &  -0.0036 &  0.0016  &   0.75  &  0.23  &  0.008 & 0.004  &  1.04  &  0.16  &  55.2  &  4.5  \\
        2011-04-29  &  5681.68  &  -0.0137 &  0.0040  &   0.75  &  0.23  &  0.012 & 0.007  &  1.65  &  0.40  &  73.1  &  6.9  \\
        2011-04-30  &  5682.61  &   0.0107 &  0.0047  &   0.70  &  0.21  &  0.022 & 0.010  &  3.18  &  0.47  &  35.2  &  4.2  \\
        2011-04-30  &  5682.70  &  -0.0023 &  0.0043  &   0.70  &  0.21  &  0.010 & 0.006  &  1.47  &  0.43  &  49.5  &  8.3  \\
        2011-04-30  &  5682.75  &  -0.0216 &  0.0008  &   0.70  &  0.21  &  0.016 & 0.005  &  2.33  &  0.08  &  79.0  &  1.0  \\
        & & & & & & & & & & &  \\ 
        2011-05-30  &  5712.54  &  -0.0240 &  0.0017  &  1.18  &  0.35  &  0.037 & 0.013  &  3.17  &  0.17  &  69.6  &  1.5  \\
        2011-05-30  &  5712.57  &  -0.0202 &  0.0059  &  1.18  &  0.35  &  0.032 & 0.017  &  2.74  &  0.59  &  68.8  &  6.2  \\
        2011-05-30  &  5712.60  &  -0.0215 &  0.0019  &  1.18  &  0.35  &  0.030 & 0.011  &  2.55  &  0.19  &  73.7  &  2.1  \\
        2011-05-30  &  5712.65  &  -0.0224 &  0.0015  &  1.18  &  0.35  &  0.031 & 0.011  &  2.63  &  0.15  &  74.2  &  1.6  \\
        2011-06-01  &  5714.58  &   0.0134 &  0.0119  &  0.64  &  0.19  &  0.023 & 0.014  &  3.59  &  1.19  &  34.0  &  9.5  \\
        2011-06-01  &  5714.60  &  -0.0097 &  0.0044  &  0.64  &  0.19  &  0.015 & 0.007  &  2.34  &  0.44  &  57.2  &  5.3  \\
        2011-06-01  &  5714.63  &  -0.0067 &  0.0015  &  0.64  &  0.19  &  0.004 & 0.002  &  0.68  &  0.15  &  93.8  &  6.2  \\
        2011-06-01  &  5714.67  &   0.0025 &  0.0019  &  0.64  &  0.19  &  0.011 & 0.005  &  1.78  &  0.19  &  40.9  &  3.0  \\
        2011-06-01  &  5714.70  &  -0.0067 &  0.0005  &  0.64  &  0.19  &  0.007 & 0.002  &  1.11  &  0.05  &  63.7  &  1.3  \\
        & & & & & & & & & & &  \\ 
        2011-08-28  &  5802.44  &  -0.0046 &  0.0025  &  1.44  &  0.43  &  0.033 & 0.013  &  2.28  &  0.25  &  50.8  &  3.1  \\
        2011-08-30  &  5804.43  &  -0.0029 &  0.0027  &  0.80  &  0.24  &  0.013 & 0.006  &  1.60  &  0.27  &  129.8 &  4.9  \\
        & & & & & & & & & & &  \\ 
        2012-04-19  &  6037.67  &  0.0013  &  0.0006  &  1.07  &  0.32  &  0.015 & 0.005  &  1.36  &  0.06  &  137.8 &  1.2  \\
        2012-04-19  &  6037.70  &  0.0025  &  0.0005  &  1.07  &  0.32  &  0.008 & 0.003  &  0.73  &  0.05  &  145.2 &  2.1  \\
        2012-04-19  &  6037.73  &  0.0103  &  0.0013  &  1.07  &  0.32  &  0.013 & 0.005  &  1.20  &  0.13  &  164.7 &  3.1  \\
        2012-04-19  &  6037.75  &  0.0023  &  0.0037  &  1.07  &  0.32  &  0.014 & 0.008  &  1.32  &  0.37  &  219.9 &  8.0  \\
        2012-04-19  &  6037.79  &  0.0032  &  0.0016  &  1.07  &  0.32  &  0.004 & 0.003  &  0.33  &  0.16  &  186.9 & 13.3  \\
        2012-04-22  &  6040.70  &  0.0047  &  0.0207  &  1.49  &  0.45  &  0.067 & 0.051  &  4.48  &  2.07  &  222.0 & 13.3  \\
        2012-04-22  &  6040.73  &  0.0476  &  0.0243  &  1.49  &  0.45  &  0.097 & 0.065  &  6.49  &  2.43  &  158.6 & 10.7  \\
        & & & & & & & & & & &  \\  
        2012-05-16  &  6064.67  &  0.0374  &  0.0033  &  0.91  &  0.27  &  0.042 & 0.015  &  4.62  &  0.33  &  162.0 &  2.0  \\
        2012-05-17  &  6065.58  &  0.0436  &  0.0058  &  1.37  &  0.41  &  0.07  &  0.071 & 0.029  &  0.58  &  163.6 &  3.2  \\
        2012-05-17  &  6065.61  &  0.0351  &  0.0024  &  1.37  &  0.41  &  0.06  &  0.056 & 0.020  &  0.24  &  164.3 &  1.7  \\
\hline

\hfill{}

\end{longtable}
\tablefoot{{\it PA} values have a multiplicity of $+n180^\circ$.}

}

\section{Discussion}
\label{dis}

\subsection{The 7 mm light curve}
\label{disr}

We have shown that what looked like a single extremely strong radio flare at 7 mm in our single-dish observations, starting during the second semester of 2011 and ending one year later, is better interpreted as the superposition of three flares delayed with relation to the respective $\gamma$-ray flares. A similar delay was found for a fourth isolated flare, observed at 7 mm on 30 November 2012, which we believe is the counterpart of the $\gamma$-ray flare of 23 September 2012.

The existence of a correlation and delays between the variabilities favours the scenario of a common origin of  $\gamma$-ray and radio flares, like that proposed by the shock-in-jet models and their generalizations \citep{mar85,hu85,mar90,mar92,ste96,tur99,sok04}. In these models, electrons are accelerated to ultra-relativistic energies in a shock, and as they propagate along the jet they emit synchrotron radiation at low frequencies producing high-energy photons by the inverse Compton process. The delay is explained if the shock is formed close to the core, where the jet is optically thick to radio frequencies, but transparent to high energies, so that photons from either the BL region or DT could produce the observed high-energy emission \citep{ale14}.
 
Other evidence confirms this interpretation. First, by comparing the single-dish  7 mm light curve with the flux density of the VLBA core at the same frequency in Figure \ref{all}, we verify that the single-dish flux density was always larger during  flares, implying that when the new components become optically thin, they have already left the core. In fact, \citet{ori13} identified  a new component in the 15 GHz VLBA images, ejected from the core in 2012 (July or October)  at a velocity of $(0.92 \pm 0.35)$ mas yr$^{-1}$; after our measured delay of 54 days, when the component  became optically thin at 7 mm,  its distance to the core was $(0.14 \pm 0.05)$ mas, being resolved by the 0.1 mas VLBA beam \citep{mar12}. 
The formation of new components associated with  $\gamma$-ray flares was also confirmed by inspection of the 7 mm VLBA images obtained by the VLBA-BU-BLAZAR programme, which also showed  the polarized flux. 



Comparing our radio light curve with those at other frequencies from the F-GAMMA, GASP, and OVRO programmes \citep{ori13,ale14} at the same  epochs, we can see a compatible behaviour. Analysing  the very well-sampled 15 GHz and 22 GHz light curves obtained by \citet{ori13}, we  can also see a sharp increase in flux density during the second semester of 2011, which seems to have occurred with a delay  of approximately 50 and 35 days, respectively, relative to our 43 GHz data.  At 86 GHz and 142 GHz,  the light curves  show  a maximum very close to the second $\gamma$-ray flare in our Figure \ref{new}, indicating that they are possibly associated.

We also verified that in the 22 GHz light curves presented by \citet{ori13} using VERA, there was no difference between the single-dish flux density and that of the VLBI  core, which can be understood considering that new components became optically thin at this frequency 90 days after the $\gamma$-ray flare, at a distance of 0.22 mas from the core, not being resolved by the $(1.5 \times 1.0)$ mas VERA beam.


\citet{ale14} presented the spectral evolution of PKS1510-089 over 80 days, from MJD 55954 (01/27/2012) to MJD 56034 (04/16/2012); in Figure \ref{lastt}a we show these spectra including our 43 GHz data. We can see that the 43 GHz data fitted the overall SED only at the first epoch, with the flux density starting to increase continuously afterwards, as expected if a new component was formed, which remained optically thick at low frequencies and started to become optically thin at 43 GHz. 
In Figure \ref{lastt}b we show the SED at earlier epochs than those presented in Figure \ref{lastt}a, with our 43 GHz data superposed to those presented in \citet{ori13}. The first epoch corresponds to the quiescent period of PKS 1510-089, with a smooth and almost flat SED. On MJD 55862 the SED  presented two different peaks, at 22 GHz and 86 GHz, with flux densities close to their maximum values, which we can interpret  as  evidence of the contribution of  two components. The low-frequency peak  corresponds to the first flare in our Figure \ref{new}, becoming optically thin 35 days after 43 GHz, as mentioned before, while the high-frequency peak  corresponds to the component formed during the second $\gamma$-ray flare, on MJD 55858.  It is the high-frequency component that  increased the flux density at 43 GHz in MJD 55917.

 \begin{figure}
        \includegraphics[width=9cm]{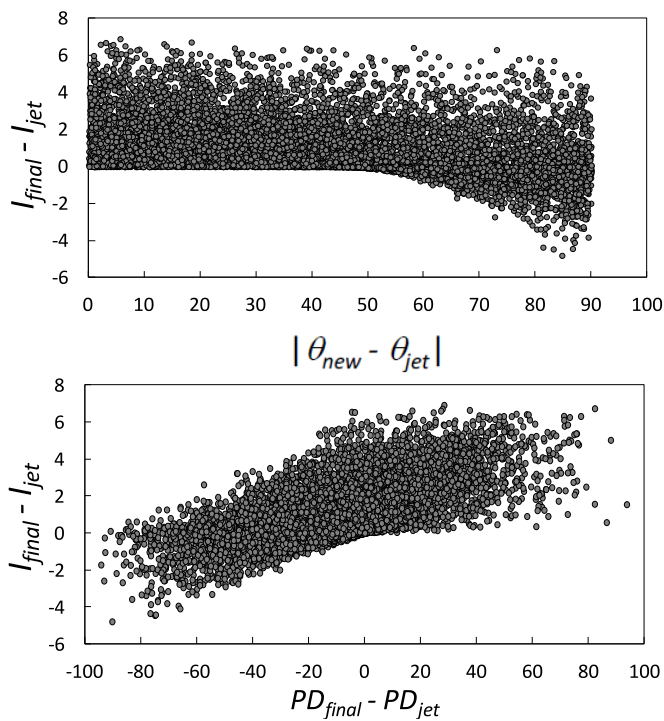}
   \centering
        \caption{Difference between polarized fluxes before  and after the formation of a new component as a function of the difference in {\it PA} of the new and old component (upper panel) and difference in {\it PD} before and after the formation of the new component (lower panel).}
        \label{Fig9}
\end{figure}

\begin{figure}
        \includegraphics[width=9cm]{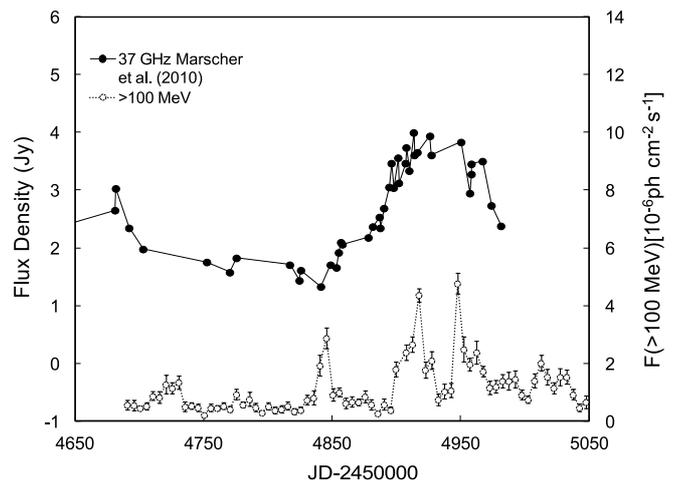}
    \centering
        \caption{Light curves of PKS 1510-089 at 37 GHz (black dots) obtained by \citet{mar10} and at $\gamma$-rays (band 3 data from Fermi/LAT, white dots). The radio light curve was shifted by 54 days, the same time delay as in  Fig. \ref{G7}.}
        \label{G7mar}
\end{figure}

A similar delay between radio and $\gamma$-ray flares seems to have occurred in 2009 \citep{mar10}, but it was not covered by our radio observations. At that epoch, three  groups of $\gamma$-rays flares were observed, separated by about 50 days, with a sharp increase in flux density at 15 GHz, 37 GHz, and 230 GHz coinciding with the last of the flares.
Taking into account the  time delay between radio and $\gamma$-ray emission, we believe that -- instead of a coincidence with the last flare -- the increase in the radio flux density could be attributed to a delayed synchrotron emission from the previous one. In Figure \ref{G7mar} we present the radio light curve obtained by \citet{mar10} shifted by the same delay adopted in Figure \ref{G7} (54 days). With this displacement in time, the beginning of the increase in radio flux density seems to be associated with the first $\gamma$-ray flare. Naturally,  flare delays in 2009 and 2011 do not need to have the same value, but as  can be seen, a  delay similar to that in 2011 could also explain the 2009 variability.


\begin{table}
 {\small
 \caption{Different value of {\it PA} in our data after their combination with data from the literature}
 \hfill{} 
 \begin{tabular}{ c c c c}
\hline\hline Date & JD & $PA$ in Fig. \ref{all} & PA in Fig. \ref{last} \\
\hline 2009-03-31   & 4922.81   & 1.2   & 181.2  \\
2009-03-31   & 4922.85   & 2.6   & 182.6   \\
2009-04-01   & 4923.66   & -36.8   & 143.2  \\
2009-04-01   & 4923.79   & -38.2   & 141.8  \\
2009-04-20   & 4942.62   & -59.2   & 120.8  \\
2009-04-20   & 4942.65   & -59.2   & 120.8  \\
2009-04-22   & 4944.63   & 5.1   & 185.1  \\
2009-04-22   & 4944.7   & 16.7   & 196.7  \\
2009-04-23   & 4945.7   & 25.5   & 205.5  \\
2009-04-23   & 4945.73   & 13.2   & 193.2 \\
2009-04-23   & 4945.76   & 12.1   & 192.1 \\
2009-23-04   & 4945.8   & 13.7   & 193.7 \\
2009-05-20   & 4972.55   & 98   & -82.0 \\
2009-05-21   & 4973.63   & 60.9   & 60.9  \\
2009-05-22   & 4974.57   & 72.4   & 72.4 \\
2009-05-22   & 4974.6   & 72.8   & 72.8  \\
2009-06-03   & 4986.57   & 136.2   & 136.2   \\
2009-06-22   & 5005.54   & 189.6   & 369.6   \\
2009-06-23   & 5006.54   & 164.2   & 344.2   \\
2009-06-23   & 5006.57   & 159.1   & 339.1   \\
2009-06-23   & 5006.6   & 142.4   & 322.4  \\
2009-06-23   & 5006.63   & 145.2   & 325.2 \\
2009-06-24   & 5007.49   & 112.4   & 292.4  \\
2009-07-13   & 5026.48   & 86.1   & 446.1  \\
2009-07-13   & 5026.52   & 90.7   & 450.7  \\
2009-07-13   & 5026.56   & 85.6   & 445.6   \\
2009-07-14   & 5027.47   & 85.1   & 445.1  \\
2009-07-14   & 5027.52   & 88.6   & 448.6  \\
2009-07-14   & 5027.56   & 89.1   & 449.1 \\
\hline
\end{tabular}}
\hfill{}
\label{PAt}
\end{table} 

\subsection{Stokes parameters of a new component}
\label{disp}

The formation of  new jet components can also explain the behaviour of the optical polarization, which shows, during flares,  variability in both {\it PA} and {\it PD}, in only one of them, or in neither of them.
To understand this behaviour, let us consider as in \citet{hol84}, the Stokes parameters for two components; in our case, before and after the formation of a new component.  We represent the  jet as a single structure before the formation of the new component, with Stokes parameters

\begin{equation}
\label{eq1}
Q_{jet} = I_{p(jet)}\cos{2 \theta_{jet}},
\end{equation}
\begin{equation}
\label{eq2}
U_{jet} = I_{p(jet)}\sin{2 \theta_{jet}},
\end{equation}
\noindent
where $\theta_{jet}$ is the {\it PA} and $I_{p(jet)}$ is the total polarized flux density before the formation of the new component. The Stokes parameters of the source after the ejection ($Q_{final},U_{final}$, and $I_{p(final)}$) can be written as

\begin{equation}
\label{eq3}
Q_{final} = Q_{jet}+Q_{new}
,\end{equation}
\label{eq4}
\begin{equation}
U_{final} = U_{jet}+U_{new}
,\end{equation}
\begin{equation}
\label{eq5}
I_{p(final)} = \sqrt{Q_{final}^{2}+U_{final}^{2}}
,\end{equation}

\noindent
where  $Q_{new}$ and $U_{new}$ are the Stokes parameters  of the new component. Solving these equations for the total polarization angle $\theta_{final}$ and for the polarized intensity $I_{p(final)}$, we have

\begin{equation}
\label{eq6}
\cos{2 \theta_{final}} = (I_{p(jet)}\cos{2 \theta_{jet}}+I_{p(new)}\cos{2 \theta_{new}})/I_{p(final)}
,\end{equation}
\begin{equation}
\label{eq7}
\sin{2 \theta_{final}} = (I_{p(jet)}\sin{2 \theta_{jet}}+I_{p(new)}\sin{2 \theta_{new}})/I_{p(final)}
,\end{equation}
\begin{equation}
\label{eq8}
I_{p(final)}^{2} = I_{p(jet)}^{2}+I_{p(new)}^{2}+2I_{p(jet)}I_{p(new)} \cos{2(\theta_{new}-\theta_{jet})}   
.\end{equation}

We  investigate the possibility that $I_{p(final)} < I_{p(jet)}$ simultaneously with a large change in {\it PA}, as detected during the $\gamma$-ray flare of April 2009:  $I_{p(jet)}=(0.18 \pm 0.07) $ mJy and $\theta_{jet}=-59^\circ \pm 2^\circ$ on 20 April, and $I_{p(final)}=(0.08 \pm 0.02)$ mJy and $\theta_{final}=+11^\circ \pm 2^\circ$ on 22 April.   
From equation \ref{eq8}, this condition is satisfied if

\begin{equation}
\label{eq9}
I_{p(new)} < -2I_{p(jet)}\cos{2(\theta_{new}-\theta_{jet})} 
\end{equation}

\noindent
or  
\begin{equation}
\label{eq10}
\mid(\theta_{new}-\theta_{jet})\mid>45^\circ. 
\end{equation}

We  used  equations \ref{eq6} to \ref{eq8} to estimate the polarimetric properties of the ejected component and found  $\theta_{new}={23\fdg 7}  ^{+2\fdg 8}_{-3\fdg 3} $ and $I_{p(new)} = (0.23\pm 0.10)$ mJy, resulting  in $(\theta_{new}-\theta_{jet})=83^\circ \pm 4^\circ $, in agreement with the requirement of equation \ref{eq10}.

Based on equations \ref{eq6} to \ref{eq8}, we also checked the incidence rate of variations in {\it PA} and  {\it PD} after the ejection of a new component, simulating different combinations between {\it PA}, {\it PD,} and total flux of the jet and of the new component. We  used a uniform distribution of random values for the intensities (between 0 and 7 mJy), {\it PA}s (between $0^\circ$ and $180^\circ$), and {\it PD}s (between 0 and 100$\%$). 
In Figure \ref{Fig9} (top) we present $(I_{p(final)}-I_{p(jet)})$ as a function of $\Delta \theta = (\theta_{new}-\theta_{jet})$ and verify  that the polarized flux density can indeed decrease after the formation of a new polarized component if equation \ref{eq10} is satisfied.
In Figure \ref{Fig9} (bottom) we present $(I_{p(final)}-I_{p(jet)})$ as a function of $({\it PD}_{final}-{\it PD}_{jet})$, which shows that if the polarized flux density decreases after the formation of the new component, ${\it PD}_{final}$ will be lower than  ${\it PD}_{jet}$. 
We also did  simulations using a normal  instead of a uniform distribution for the total flux densities, {\it PA}s, and {\it PD}s, and the results did not change significantly, indicating that the {\it PA} combination is the most important variable to obtain ${\it PA}_{final}-{\it PA}_{jet}$.

The existence of both large and small changes in {\it PD} with the same probability, also explains $\gamma$-ray flares without any corresponding change in {\it PA} or {\it PD}, as reported  by \citet{pal15} for an extremely bright  $\gamma$-ray flare observed during 2014 in the FRSQ 3C279. Those authors also pointed out that such behaviour can be  evidence of a superposition of multiple components. They are also in agreement with the first year results of the RoboPol (Robotic Polarimetric in Crete), a monitoring programme of an unbiased sample of  $\gamma$-ray bright blazars, which suggest that the highest amplitude $\gamma$-ray flares are correlated with rotations in $PA$ \citep{bli15}.

The results of our simulations are different from those reported by \citet{kie13} because they assumed variability due to random effects, used a large number of components (30), and attributed random values for {\it Q} and {\it U} instead of intensity, $PD,$ and $PA$, as in our case. The analysis was carried out based on a sample with a regular small time interval of three days and, as was also pointed by the authors, poorer sampled light curves can introduce large artificial rotations.

\subsection{Optical {\it PA} variability }

  \begin{figure*}
        \center
        \includegraphics[width=16 cm]{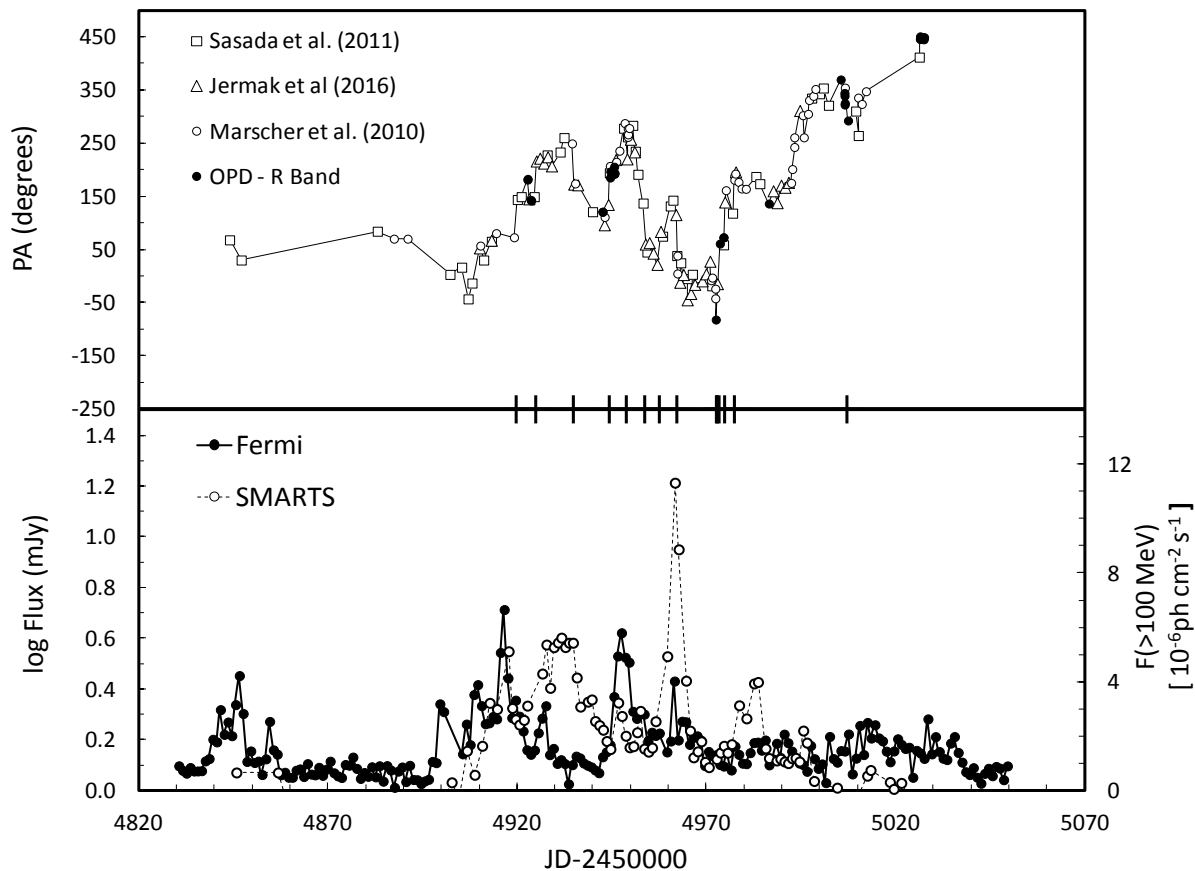}
        \caption{Top: {\it PA} variability combining the results obtained in this work with those of \citet{mar10}, \citet{sas11}, and \citet{jem16}.  Bottom:  SMARTS $R$ band and Fermi/LAT light curves. Vertical marks in the time axis represent the epochs in which rotations of more than $50^\circ$ were detected in intervals of  one day.}
        \label{last}
  \end{figure*}


As pointed out by \citet{kie16}, the existence of  gaps in the observations compromises the solution of the $180^\circ$ multiplicity {\it PA}. Attempting to better constrain the {\it PA} variability, we combined the polarimetric data reported by \citet{mar10}, \citet{sas11}, \citet{jem16}, and our own data, and applied as a criterion to keep the smallest variation in {\it PA} during successive epochs, allowing rotation in both clockwise and anticlockwise senses. 

The resulting {\it PA}s for our data are presented in Table 3 and the behaviour of {\it PA} including all the data are shown in Figure \ref{last}. We can see that the largest jumps in {\it PA} coincided with the occurrence of optical and/or $\gamma$-ray flares, after which {\it PA} tended to return to its previous value. 
The maximum rotation during this time interval was less than $400^\circ$, in agreement with \citet{sas11} and \citet{jem16}. Figure \ref{last}, with all data combined, shows in more detail the oscillations associated with the occurrence of optical and/or $\gamma$-ray flares. 


%

\section{Conclusions}
\label{conclusion}

In this paper we presented  7 mm single-dish observations of PKS 1510-089 covering the period January 2011 to September 2013 and $R$-band polarimetric observations at several epochs between March 2009 and May 2012.
The long coverage of the 7 mm light curve allowed us to correlate four $\gamma$ rays with radio flares and measure a delay of about 54 days between them, with the $\gamma$-ray emission occurring first. This result shows that sometimes the simultaneity between flares at the two frequencies can be a consequence of coincidence between the delay timescales and the rate of occurrence of the flares. 

A comparison between the measured single-dish 7 mm flux density and the VLBI core emission at the same wavelength shows that the components responsible for the flares had already left the core when they become optically thin, although this was not the case at lower frequencies. These results are a consequence of the apparent velocity of the components, the time delay between gamma and radio flares, and the resolution of the different VLBI arrays used in the observations.

We detected a large rotation ($65^\circ$) in {\it R}-band {\it PA} simultaneously with the start of a $\gamma$-ray flare in April 2009. Combining our observations with data from the literature, we verified that this simultaneity also occurred in other $\gamma$-ray flares.
We showed that the behaviour of the polarization variability, which included sudden large variations in {\it PA} and {\it PD}, in only one of them, or in neither of them during flares can be explained by the superposition of the jet emission and that of a new component.

\begin{acknowledgements}
We are grateful to the Brazilian research agencies FAPESP and CNPq for financial support (FAPESP Projects: 2008/11382-3 and
 2014/07460-0).
This study makes use of 43 GHz VLBA data from the VLBA-BU Blazar Monitoring Program (VLBA-BU-BLAZAR;
 http://www.bu.edu/blazars/VLBAproject.html), funded by NASA through the Fermi Guest Investigator Program. The VLBA is an instrument of
  the National Radio Astronomy Observatory. The National Radio Astronomy Observatory is a facility of the National Science Foundation
   operated by Associated Universities, Inc. This paper has made use of up-to-date SMARTS optical/near-infrared light curves that are
    available at www.astro.yale.edu/smarts/glast/home.php. This research has made use of data from the MOJAVE database that is maintained
         by the MOJAVE team (Lister et al., 2009, AJ, 137, 3718). 
\end{acknowledgements}

%
%

\end{document}